\documentclass[twocolumn,showpacs,preprintnumbers,amsmath,amssymb]{revtex4}

\usepackage{graphicx}
\usepackage{dcolumn}
\usepackage{bm}

\begin{document}

\preprint{APS/123-QED}

\title{Drift velocity peak and negative differential mobility in high field transport in graphene nanoribbons explained by numerical simulations}


\author{Alessandro Betti}
\affiliation{Dipartimento di Ingegneria dell'Informazione: Elettronica, Informatica, Telecomunicazioni, \\Universit\`a di Pisa, Via Caruso 16, 56122 Pisa, Italy.}
\author{Gianluca Fiori}
\affiliation{Dipartimento di Ingegneria dell'Informazione: Elettronica, Informatica, Telecomunicazioni, \\Universit\`a di Pisa, Via Caruso 16, 56122 Pisa, Italy.}
\author{Giuseppe Iannaccone}
\affiliation{Dipartimento di Ingegneria dell'Informazione: Elettronica, Informatica, Telecomunicazioni, \\Universit\`a di Pisa, Via Caruso 16, 56122 Pisa, Italy.}


\begin{abstract}

We present numerical simulations of high field transport in both
suspended and deposited armchair graphene nanoribbon (A-GNR) on HfO$_2$ 
substrate. Drift velocity in suspended GNR does not saturate 
at high electric field ($F$), but rather decreases, showing a maximum for 
$F\approx 10$~kV/cm. 
Deposition on HfO$_2$ strongly degrades the drift velocity 
by up to a factor of $\approx 10$ with respect to suspended GNRs in the 
low-field regime, whereas at high fields drift velocity approaches the 
intrinsic value expected in suspended GNRs. 
Even in the assumption of perfect edges, the obtained mobility is far 
behind what expected in two-dimensional graphene, and is further reduced by 
surface optical phonons. 

\end{abstract}

\pacs{73.63.-b,73.50.Dn,72.80.Vp,63.22.-m}
\maketitle

\newpage

Assessing the potential of Graphene Nanoribbons (GNRs) for future 
electronic applications requires full understanding of both 
quasi-equilibrium and far-from-equilibrium transport 
mechanisms~\cite{Wang,ABettiTED2}. 
However experimental 
low-field (LF) mobility in 1~nm-wide GNRs can be as low as 
100~cm$^2$/Vs and is limited by edge disorder~\cite{Wang,Yang}. 
Furthermore, we know that achieving ideally smooth edges is not enough:
full-band (FB) modeling shows that LF mobility due to only 
acoustic (AC) phonon scattering at
low fields is close to 500~cm$^2$/Vs for 1 nm-wide GNRs~\cite{Bettiremote}. 
Most importantly, nanoscale transistors do not operate in the LF 
mobility limit.
Therefore, simulation of far-from-the-equilibrium transport conditions is 
required to understand
achievable device performance. Whereas at low field carrier scattering 
is mainly due to low-energy 
intravalley acoustic phonons~\cite{Bettiremote}, at high electric field 
scattering is dominated by 
optical phonon emission (EM), which becomes relevant when electrons 
gain enough 
energy to emit optical phonons. High-field steady-state transport can be 
simulated by solving the Boltzmann transport 
equation (BTE) through the single-particle Monte Carlo (MC) method. 
When dealing with 1D systems, however, particular attention has to be paid in 
accurately describing the energy dispersion relation due to quantum 
lateral confinement: for each GNR subband there is a parabolic behavior close 
the subband minimum, and then the typical graphene quasi-linear behavior 
already for relatively small wavevector values, corresponding to a velocity $\approx 8 \times 10^5$~m/s.
Some authors have considered phonon confinement and multisubband transport, 
focusing on quantum wires with the effective mass 
approximation~\cite{Briggs,Mickevicius_high}. 
For GNRs, a multisubband MC approach has been followed in~\cite{Zeng} and BTE has been solved in a 
deterministic way at criogenic temperatures in~\cite{Huang}. 
Bresciani {\em et al.}~\cite{Bresciani} have used a 2D model, 
which is not fully adequate 
for sub-10~nm GNRs, where size effects are indeed 
relevant. 

In this work, we adopt a steady-state single-particle full band MC approach 
accounting for carrier degeneracy \cite{Lugli},  which 
has a significant
effect for materials with a small density of states as graphene.
Scattering rates are obtained within the Deformation Potential Approximation 
(DPA) from phonon dispersions described by means of the fourth-nearest-neighbour 
force-constant approach (4NNFC)~\cite{Saito} and a p$_z$ 
tight-binding Hamiltonian for the electronic structure. 
We consider in-plane longitudinal acoustic and optical (LA and LO), 
transversal optical (TO) and 
surface optical (SO) phonons. In each 
subband, the rates are computed on a 2000-point grid in the $k_x$-space 
(due to simmetry only longitudinal electron wavevectors 
$k_x >0$ have been taken into account), 
considering energy up to 1.5~eV above the bottom of the 
first subband and including up to 18 subbands: this ensures accurate 
results even for strong longitudinal electric field 
$F \leq 3 \times 10^2$~kV/m and for all the considered 
GNR widths ($W \leq$~10~nm). 
Due to Van Hove singularities in the 1D rates, the self-scattering 
method~\cite{Jacoboni} is inefficient, so that we have 
adopted the MC procedure described in Ref.~\cite{Jacoboni} and 
extended to quasi-1D systems~\cite{EPAPS}. 
The whole story of an electron has a duration $T =\sum_i \Delta t_i $ 
($T \leq (10-100)$~$\mu s$ depending on $W$ and $F$), 
where $\Delta t_i= t_{i+1}-t_i$ is the $i$-th time step and $t_i$ is 
the $i$-th sampled time. At each $t_i$, the wavevector 
${\bf k}=(k_x,k_{y\eta})$, where $k_{y\eta}$ is the transverse quantized 
electron wavevector~\cite{Bettiremote}, is computed and the 
average value of quantity $\mbox{X}$ (either the drift 
velocity $v({\bf k})= 1/\hbar \cdot \partial E({\bf k}) /\partial k_x$ 
or the energy $E({\bf k})$) is evaluated according to~\cite{Jacoboni}:
$ \langle \mbox{X} \rangle_t = \frac{1}{T} \int_0^T \mbox{X}\left[k_x(t') \right] dt'= 
\frac{1}{T}\sum_{t_i} \int_{t_i}^{t_{i+1}} \mbox{X} \left[k_x(t')\right] dt' \, .$
The electronic temperature $T_{el}$ under an applied homogeneous  field $F$ is 
computed by equating $\langle v(k_x)^2 \rangle$ with 
the mean squared velocity at equilibrium~\cite{EPAPS}. 
For $F=0$, $T_{el}$ is equal to the lattice room temperature $T_{lat}$, 
whereas for $F>0$, $T_{el} > T_{lat}$. 
Once obtained $T_{el}$, the distribution function 
is updated accordingly at each $t_i$ and final states for scattering are 
filled obeying to the Pauli exclusion principle (PEP). 
\begin{figure} [tbp]
\begin{center}
\includegraphics[width=7.5cm]{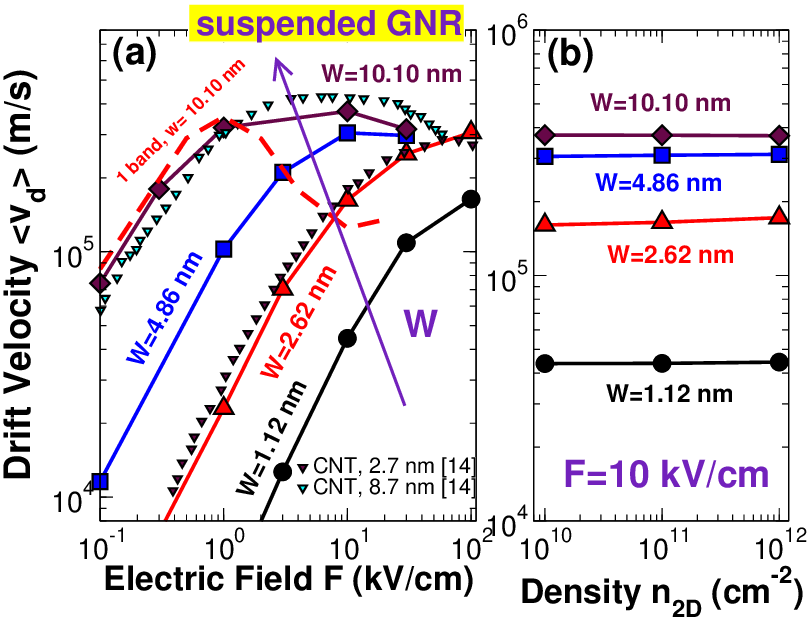}
\end{center}
\vspace{-0.6cm}
\caption{(Color online) (a) Average drift velocity 
$\langle v_{d} \rangle$ as a 
function of $F$ ($n_{2D}=$~10$^{12}$~cm$^{-2}$) and (b) as a function of 
$n_{2D}$ for different $W$ of suspended GNRs. 
In (a) MC results for zigzag 
CNTs~\cite{Pennington} with sub-10~nm circumference (2.7~nm and 8.7~nm) are 
also reported. In (b) $F=$~10~kV/cm.}
\label{fig:fig1}
\end{figure} 

The computed average drift velocity $\langle v_{d} \rangle $ limited by 
intrinsic phonons is plotted in Fig.~\ref{fig:fig1}a as a function of $F$ 
for suspended sub-10~nm GNRs and for a carrier density 
$n_{2D}=$~10$^{12}$~cm$^{-2}$. 
$\langle v_{d} \rangle $ strongly varies within the considered $F$ interval 
and a maximum appears at high electric field ($F=F_{th}$). 
For $F < F_{th}$, $\langle v_{d} \rangle$ can be fitted 
by means of the Caughey-Thomas model, i.e. 
$ \langle v_{d}(F) \rangle = (\mu_{in,0} F)/[1+(\mu_{in,0} F/v_p)^{\gamma}]^{1/\gamma}$, 
where $\gamma \approx (1.3 \div 2)$, $\mu_{in,0}$ is the intrinsic LF 
mobility and $v_p$ is the peak velocity, which ranges from 
$ 2 \times 10^5$ to $4\times 10^5$~m/s, depending on $W$ 
(Table~\ref{tab:table}).
$\langle v_{d} \rangle$ does not saturate with $F$, analogously to what 
has been observed 
in the case of zigzag CNTs ~\cite{Pennington,PerebeinosPRL}.
As shown in Fig.~\ref{fig:fig1}a,  $\langle v_{d} \rangle$ slightly decreases 
for increasing $F$~\cite{Pennington}. However, in constrast 
to~\cite{Pennington}, 
the negative differential mobility (NDM) 
cannot be explained by the increased number of populated 
states in the second subband with smaller $ v ({\bf k})$. 
Indeed, as in~\cite{PerebeinosPRL}, we have verified for $W=$~10~nm that 
such effect is  much more pronounced if we fictitiously limit transport to only one subband 
(red dashed line in Fig.~\ref{fig:fig1}a). 
\begin{table}[htbp]

\caption{Caughey-Thomas parameters $\mu_{in,0}$ (cm$^2$/Vs), 
$v_p$ (m/s) and $\gamma$ for 
different $W$ (nm), obtained through the fitting with MC data.}
\begin{center}
\begin{tabular}{l  l  l  l  l  l  l  l}
\hline
$W$ &\, $\mu_{in,0}$ & \, $v_p$ & $ \gamma$ & $W$ & \, $\mu_{in,0}$ & \, $v_p$ & $ \gamma$ \\
\hline  
1.12 &\, 310 & \, $2.2 \times 10^5$ & 1.7 & 4.86 & \, 12000 & \, $3.3 \times 10^5$ & 1.4 \\
\hline
2.62 &\, 2700 & \, $3.2 \times 10^5$ & 1.3 & 10.10 & \, 80000 & \, $3.7 \times 10^5$ & 2 \\
\hline
\end{tabular}
\end{center}
\label{tab:table}
\end{table} 
The velocity peak can be explained by the combined effect of the quasi-linear 
dispersion relation in GNRs and the 
strong increase of optical EM at high $F$, which increases the occurrence of 
backscattering events
therefore reducing the average velocity. If we follow the story of a 
single electron, 
we can see that as $F$ increases, instantaneous electron velocity cannot 
increase beyond the 
limit imposed by the dispersion relation: any backscattering event will 
invert the instantaneous 
velocity sign and therefore reduce the average velocity. Therefore, in the 
absence of optical 
phonon EM, the drift velocity $\langle v_d \rangle$ would saturate to 
about $8 \times 10^5$~m/s. 
The onset of optical phonon EM makes  $\langle v_d \rangle$ peak at a 
fraction of that value and then decrease with $F$. 
We believe that the very same mechanism explains also the NDM in zigzag 
CNTs, as well as in graphene~\cite{Li}, 
even if it has not been proposed before~\cite{Pennington,PerebeinosPRL}.
One can also see that the peak velocity $v_p$ increases with $W$. 
Indeed, if $\hbar \omega $ is the optical phonon energy 
($\approx 160$~meV for LA mode), the current $J$ can be estimated as 
$J= \frac{4 e}{h} \hbar \omega \frac{E_F}{\pi \hbar v_F}$~\cite{Freitag}, 
where $v_F$ is the Fermi velocity. 
Since $v_p= J/(e n_{2D})$ and the Fermi wavevector $k_F= \pi n_{2D} W$, 
we obtain $v_p= \omega \times W/\pi \propto W$. 

The threshold field $F_{th}$ strongly decreases 
with $W$, because of the increased mean free path 
$\langle L_{\bf k} \rangle$, which allows electrons to gain energy
required to emit optical phonons at lower $F$. 
$F_{th}$ can be roughly extimated by imposing the 
cutoff energy $\hbar \omega $ for optical EM equal to the mean kinetic
energy gained between two scattering events, 
i.e. $q F_{th} \langle L_{\bf k} \rangle $. 
Since by increasing $W$ from 1 to 10~nm, 
$\langle L_{\bf k} \rangle$ increases from $\approx 6$~nm to 
1~$\mu$m~\cite{Bettiremote}, $F_{th}$ 
decreases from $\approx 2.7 \times 10^2$~kV/cm 
to $\approx 1.6$~kV/cm. 
We remark also that the obtained values for $\langle v_{in} \rangle $ are 
in agreement with those found for zigzag CNTs with a circumference 
comparable with the considered $W$, both in linear and in non-linear 
regimes~\cite{Pennington}. 
In Fig.~\ref{fig:fig1}b, $\langle v_{d} \rangle$ 
is plotted as a function of $n_{2D}$ for $F=$~10~kV/cm: 
$\langle v_{d} \rangle$ does not depend on $n_{2D}$ even in the degenerate 
regime ($n_{2D}=$~10$^{12}$~cm$^{-2}$), 
where PEP limits up to the 50\% of the scattering events.  
\begin{figure} [tbp]
\begin{center}
\includegraphics[width=7.5cm]{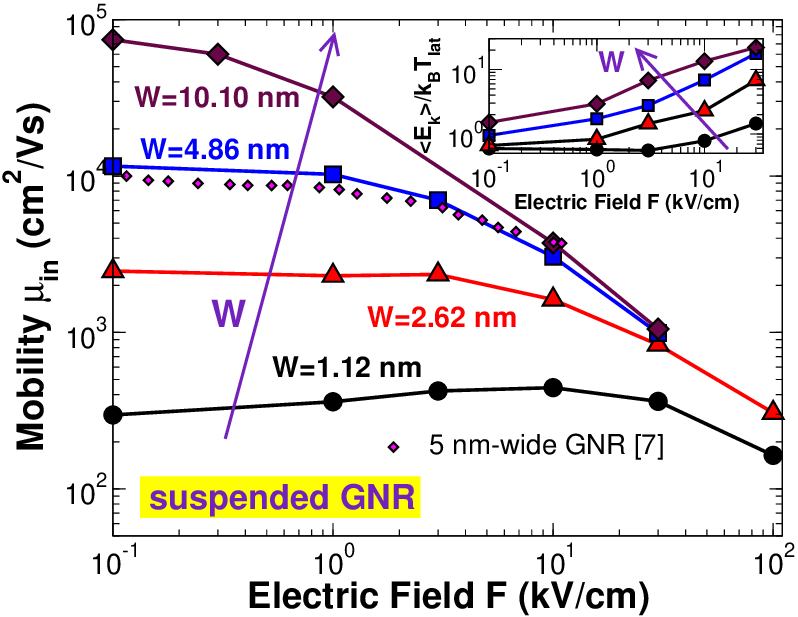}
\end{center}
\vspace{-0.6cm}
\caption{(Color online) (a) Intrinsic mobility $\mu_{in}$ 
and (Inset) average electron energy 
as a function of $F$. Results for 
GNRs with $W=$~5~nm~\cite{Zeng} are also shown in (a). 
$n_{2D}=$~10$^{12}$~cm$^{-2}$.}
\label{fig:fig2}
\end{figure} 
In Fig.~\ref{fig:fig2}, we show the intrinsic mobility 
$\mu_{in}\equiv \langle v_{d} \rangle / F$ as a 
function of $F$ for $n_{2D}=$~10$^{12}$~cm$^{-2}$. Since in the linear 
transport regime $ v_{d} \propto F$, $\mu_{in}$ is constant for 
$F \leq F_{th}$ and decreases above $F_{th}$ as 
$\mu_{in} \propto 1/F^{\alpha}$ with 
$\alpha > 1$. 
In addition, the narrower the ribbons, the stronger the suppression due to 
lateral confinement. 
Results are in good agreement with a multi-subband model for 
$W \approx 5$~nm~\cite{Zeng}. 
In the inset of Fig.~\ref{fig:fig2}, the average kinetic electron energy 
$\langle E_{\bf k} \rangle$ in units of $k_B T_{lat}$ is shown 
as a function of $F$ for $n_{2D}=$~10$^{12}$~cm$^{-2}$. 
While for low $F\approx$~0.1~kV/cm electrons tend to remain 
near the first conduction subband edge and 
$\langle E_{\bf k} \rangle \approx 1/2 k_B T_{lat}$ for the narrowest ribbons, 
for high field 
$F\geq 10^2$~kV/m, $\langle E_{\bf k} \rangle \gg k T_{lat}$ and higher 
energy states are occupied. 
Note also that $\langle E_{\bf k} \rangle$ increases 
with $W$, since subbands become closer, allowing 
electrons to populate higher subbands.   
\begin{figure} [tbp]
\begin{center}
\includegraphics[width=7.5cm]{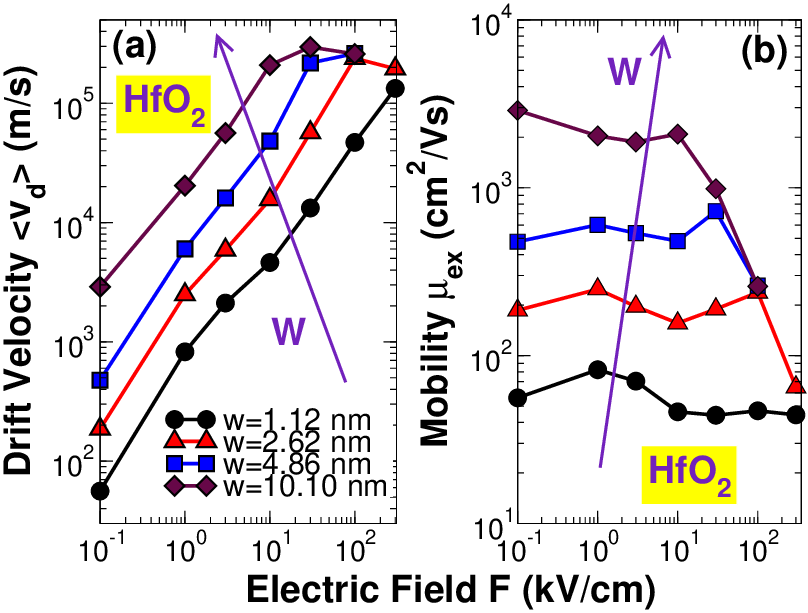}
\end{center}
\vspace{-0.6cm}
\caption{(Color online) (a) 
$\langle v_d \rangle$ and (b) mobility $\mu_{ex}$ as a function of 
$F$ for GNR on HfO$_2$. $n_{2D}=$~10$^{12}$~cm$^{-2}$.}
\label{fig:fig3}
\end{figure}
 
In Fig.~\ref{fig:fig3}a and~\ref{fig:fig3}b 
the average drift velocity $\langle v_d \rangle $ and the mobility 
$\mu_{ex}=\langle v_d \rangle / F $ are shown, respectively, 
as a function of $F$ for GNRs deposited on 
HfO$_2$~\cite{Bettiremote} including 
the effect of SO phonons ($n_{2D}=$~10$^{12}$~cm$^{-2}$), 
through the first SO(1) and second SO(2) modes 
($\hbar \omega_{SO}=$~12.4~meV for SO(1) mode). 
With respect to the suspended GNR case, 
$\langle v_{d} \rangle$, as well as mobility in GNR on HfO$_2$
are almost one order of magnitude smaller in the LF regime, 
while similar values are obtained for high field,
as already noted for graphene on HfO$_2$~\cite{Li}.
As in graphene~\cite{Li}, deposition on HfO$_2$ leads 
to an extension of the linear region to fields up to a 
factor~10 larger than those corresponding to suspended GNRs. 
For narrow ribbons, $\langle v_d \rangle$ 
does not saturate even for $F=3 \times 10^2$~kV/cm.

\begin{figure} [tbp]
\begin{center}
\includegraphics[width=7.5cm]{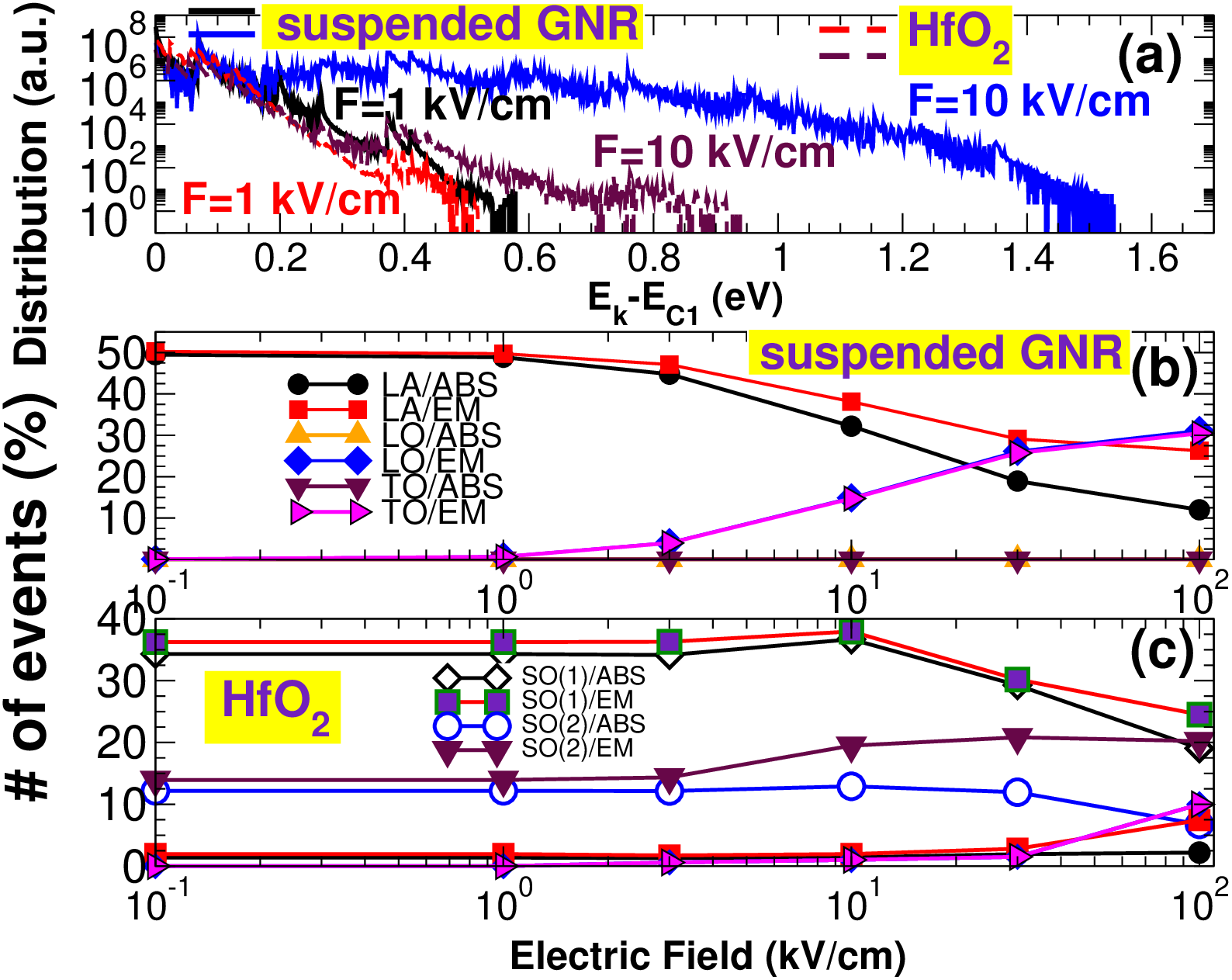}
\end{center}
\vspace{-0.6cm}
\caption{(Color online) (a) Energy distribution function for 
$F=$~1 and 10~kV/cm for both suspended GNR (continuous lines) 
and GNR deposited on HfO$_2$ (dashed lines). 
Fraction of different scattering events 
as a function of $F$ for (b) suspended GNR (LA, LO and TO phonons) 
and (c) GNR deposited on HfO$_2$ (LA, LO, TO and SO phonons). 
$W=$~4.86~nm and $n_{2D}=$~10$^{12}$~cm$^{-2}$.}
\label{fig:fig4}
\end{figure}

In order to understand the different behaviour in suspended 
and deposited GNR on HfO$_2$,
in Fig.~\ref{fig:fig4}a we show the distribution function $G$ 
as a function of $(E_{\bf k} - E_{C1})$, where $E_{C1}$ is the first 
subband edge, for $F=$~1 and 10~kV/cm ($W=4.86$~nm) 
for GNRs, both suspended and deposited on HfO$_2$. 
At low fields, $G$ decreases rapidly with energy in both cases, 
showing, unlike graphene, sharp peaks due to intersubband scattering. 
At high fields instead, electrons are excited up to energies 
close or above 1~eV, increasing $T_{el}$ (inset of Fig.~\ref{fig:fig2}). 
As expected, deposition on HfO$_2$ 
leads to a shorter high energy tail compared to that in intrinsic GNR 
(Fig.~\ref{fig:fig4}a) due 
to the introduction of an additional channel for energy and momentum 
relaxation which increases total scattering rate and pushes electrons down to 
lower average electron energy. 
As for graphene on HfO$_2$~\cite{Li}, the rough absence of degradation 
of $\langle v_d \rangle$ at high $F$ can be explained by the reduced 
population in the high energy tail, i.e. by a smaller amount of 
electrons in the nonlinear 
dispersion energy region where the band velocity is smaller, 
which more than counterbalances the decrease of $\langle v_d \rangle$ 
resulting from the increased total 
scattering rate when depositing GNR on substrate. 
In order to explain the increase of the $F$ interval where 
$\langle v_d \rangle$ 
shows a linear behavior in the GNR on HfO$_2$ case, in Fig.~\ref{fig:fig4}b 
and~\ref{fig:fig4}c 
we show the relative ratio of scattering events for the different mechanisms 
as a function of $F$ for 5~nm-wide suspended GNR and GNR on 
HfO$_2$, respectively. For intrinsic GNR and $F\leq 1$~kV/cm, 
the main scattering events 
involve absorption (ABS) and EM of LA phonons (Fig.~\ref{fig:fig4}b). 
For $F\geq 10$~kV/cm, optical phonon EM (of both LO and TO phonons) 
becomes predominant, increasing up to 30\% of the total number of events. 
For GNR on HfO$_2$ instead, scattering involving SO(1) phonons 
happens very often already for $F=1$~kV/cm whereas rates of LA, LO 
and TO phonons are limited to few percents even for high fields, as can 
be seen in Fig.~\ref{fig:fig4}c. In particular, the observed 
large absorption rate of 
SO(1) phonons (Fig.~\ref{fig:fig4}c), which is associated to the high 
Bose-Einstein occupation factor, appears to 
counterbalance 
SO(1) emission even at high $F$, and is the 
responsible of the extension of the linear region up to fields of 
10~kV/cm or above, depending on $W$.
  

In conclusion, we have performed a FB investigation of 
the dependence of drift velocity and mobility on the electric fields in GNRs. 
Suspended GNRs exhibit a drift velocity peak and then a NDM for 
large electric field, as also observed in zigzag CNTs. 
This property is due to the combined
effect of quasi-linear dispersion relation and the emission of 
optical phonons. 
In particular, the maxima occur for a threshold field 
$F_{th} \propto 1/\langle L_{\bf k} \rangle \propto 1/W$. 
Depositing GNR on HfO$_2$ substrate strongly 
degrades $\langle v_{d} \rangle$ at low field
by a factor $\approx 10$, whereas at 
high fields no degradation is observed~\cite{Li}. Furthermore, 
in deposited GNRs velocity saturation and peak are shifted at 
higher $F$, due to the compensation of SO absorption and EM mechanisms.

Authors gratefully acknowledge support from the EU FP7 Project 
NANOSIL (n. 216171), GRAND (n. 215752) grants, 
and by the MIUR-PRIN project GRANFET (Prot. 2008S2CLJ9) via 
the IUNET consortium.

\end{document}